\documentclass[aps,reprint,prc,showpacs,preprintnumbers,amsmath,amssymb,floatfix]{revtex4-2}
\usepackage{graphicx}
\usepackage{dcolumn}
\usepackage{amsmath}
\usepackage{times}
\usepackage{mathtools}
\usepackage{placeins}
\usepackage{booktabs}
\usepackage{multirow}
\usepackage{makecell}
\usepackage{xcolor}
\usepackage{cleveref}
\usepackage{enumitem}
\usepackage{verbatim}

\begin{document}

\title{Light and Strange Baryons in Medium}

\author{T.~Massimino}
\author{T.~Kl\"ahn }
\email{thomas.klaehn@csulb.edu}
\author{Z.~Papp}
\email{zoltan.papp@csulb.edu}
\affiliation{ 
California State University Long Beach, Long Beach, California, USA}

\date{\today}

\begin{abstract}
We solve the Goldstone-boson-exchange (GBE) relativistic constituent quark model of light and strange baryons by using the Faddeev approach.
The model reproduces the vacuum mass spectrum of light and strange baryons below $2$ GeV reasonably well.
To test the sensitivity of the model to possible medium-induced effects, we vary the masses of the constituent quarks and the exchange bosons, 
the confinement strength, and the quark-meson coupling constant. 
In a parametric study, we consider a set of power law scaling relations for these parameters, including some motivated by constituent quark level current algebra relations.
We find 
that the baryon spectrum is most sensitive to the quark-meson coupling constant,
and generally observe
a decrease in baryon mass with decreasing constituent quark mass.
We qualitatively estimate the impact of these mass shifts on ideal-gas baryon yields and yield ratios. 
Absolute yields can change significantly already for mass shifts of a few $10$~MeV, whereas yield ratios are strongly modified only when the compared baryons have different constituent-quark-mass dependence.

\end{abstract}

\pacs{11.30.Rd, 12.39.Pn, 25.75.Nq, 21.45.+v}

\maketitle

\section{Introduction}
When embedded in a medium at extreme density or temperature, baryon masses are expected to change.
For instance, the saturation properties of nuclear matter imply an effective or screened nucleon mass $M_N^*$ due to a large attractive scalar self-energy $\Sigma_s$~\cite{SerotWalecka97}. 
Already at saturation density $n_0\approx 0.15$ fm$^{-3}$, it is significantly reduced in comparison to the vacuum nucleon mass $M_N$.
Typical relativistic mean-field (RMF) and covariant density-functional approaches give a reduced Dirac mass at saturation of order $M_{N,0}^*\approx(0.55-0.70)M_N$~\cite{Dutra:2014qga,Niksic:2011sg}; see also \cite{Oertel:2016bki} for a review of dense matter equations of state based on RMF models.
Relativistic mean-field and nucleon potential models tie this mass reduction essentially to meson exchange between nucleons. 
More fundamentally, the constituent quark mass scale underlying the low-energy baryon mass spectrum in QCD is associated with chiral symmetry breaking~\cite{Roberts:2016vyn}.
Consequently, a partial restoration of chiral symmetry in hot or dense matter is expected to modify in-medium hadron properties, including the baryon spectrum~\cite{Pisarski:1983ms,Fukushima:2010bq}.
The connection between dynamical chiral symmetry breaking (D$\chi$SB) 
and hadron masses is particularly relevant near the QCD crossover temperature, 
$T_{\rm QCD}\approx(155\pm 5)$~MeV~\cite{HotQCD:2018pds},
where lattice QCD and Dyson--Schwinger studies indicate a smooth crossover from hadron dominated and deconfined quark--gluon degrees of freedom accompanied by a chiral transition, signaled by pronounced susceptibility peaks in close vicinity to  $T_{\rm QCD}$~\cite{Bazavov:2011nk,Fischer:2018sdj,Roberts:2000aa}.

Hadron-resonance gas (HRG) models are commonly used to interpret thermodynamic observables in terms of an ensemble of thermal, (nearly) non-interacting hadrons with constant vacuum hadron masses~\cite{Andronic:2017pug}. 
While this is a useful and successful approximation, it is still a model assumption. 
If the dynamical mass of the constituent quarks changes significantly, we expect a change of the hadron spectrum as well. 

A self-consistent computation of the full baryon spectrum from QCD remains difficult.
Relativistic mean-field models describe medium effects in terms of baryonic self-energies, motivated by meson exchanges which induce large vector and scalar mean-fields. 
Although attempts to account for medium dependent baryon masses exist (e.g. in a mean field approach~\cite{Morita:2017hgr}), 
thermodynamic models often prescribe the baryon spectrum rather than deriving it as three-quark bound states with underlying chiral dynamics~\cite{Karsch:2003vd}.
Complementary phase-shift approaches reduce the dependence on an assumed resonance spectrum by encoding interactions through scattering information~\cite{Andronic:2018qqt}.  

In this work, instead of aiming at a full thermodynamic description of three-quark bound states, we address the more limited question of how the baryon spectrum reacts to chiral restoration in the quark sector.
To do so, we solve the relativistic Faddeev equations within a potential model~\cite{papp1999treatment,day2012treatment,PhysRevC.62.044004} that reproduces the vacuum spectrum of light and strange baryons below 2~GeV.
In vacuum, this Goldstone-boson-exchange chiral quark model 
\cite{glozman1996spectrum,glozman1996light,glozman1998effective,glozman1998unified} 
reproduces the correct level ordering for the light and strange baryons without 
requiring different sets of parameters for each spectrum. 
Sections \ref{sec2}, \ref{SEC:Faddeev}, and \ref{SEC:LightAndStrangeBaryons} review the model and provide the vacuum parameterization.

To estimate the impact of chiral symmetry restoration we vary the constituent quark masses. 
This keeps the interaction structure fixed but at the same time enables us to investigate the role of the dynamical mass scale.
To account for the medium dependence of the underlying potential itself, we perform a systematic parameter study in which we employ different scaling relations between quark masses and the exchange-boson masses, meson-quark couplings, and confinement strength.
We do so to identify which model pieces affect the baryon spectrum most strongly.
As a general trend we find that most of the considered parameterizations predict decreasing baryon masses with decreasing constituent quark masses.
The strongest sensitivity arises from the scaling of the quark-meson coupling and the confinement strength.
The scaling relations are introduced in Section~\ref{sec:scaling}, their effect on the baryon spectrum is discussed in Section~\ref{sec:results}.

While different variations of scaling schemes show quantitatively significant variations with respect to the constituent quark mass dependence and certainly probe the limits of applicability of our model, the qualitative trend is a reduction of the baryon mass with decreasing constituent quark mass.
In Section~\ref{SEC:Thermo} we illustrate how chiral symmetry restoration affects particle yields and yield ratios in comparison to a model with constant, medium-independent, vacuum baryon masses.
Mainly, differences in yield ratios are expected for baryons only with different constituent quark mass dependence, whereas the actual yields will always differ from the vacuum mass yields exponentially with the mass defect.
The main results of the paper are summarized in Section~\ref{SEC:SUMMARY}.

\section{Goldstone-boson-exchange model for baryons}
\label{sec2}

We consider a three-particle Hamiltonian 
\begin{equation}\label{hamilton}
H=H^{(0)}+v_{1}+v_{2}+v_{3},
\end{equation}
where $H^{(0)}$ is the kinetic-energy operator and $v_{\alpha}$,  with $\alpha=1,2,3$, are the 
mutual interactions of the quarks.
We represent it through the usual configuration-space
Jacobi coordinates: e.g.\ ${\bf \xi}_{1}$ is the coordinate between particles $2$ and $3$ and ${\bf \eta}_{1}$ is 
the coordinate between the center of mass of the pair $(2,3)$ and particle $1$. 

The kinetic energy operator is given in the relativistic form
\begin{equation}
H^{(0)}=\sum_{i=1}^{3}\sqrt{k_{i}^{2}+m_{i}^{2}},
\label{relh0}
\end{equation}
where $m_{i}$ are the quark masses and ${\bf k}_{i}$ are the three-momenta of the quarks in the rest frame, 
where the total three-momentum
${\bf P} = \sum_{i=1}^{3} {\bf k}_{i} =0$. 

The quark-quark interaction is a long range confinement potential plus a Goldstone boson exchange short-range hyperfine interaction 
\cite{glozman1996spectrum,glozman1996light,glozman1998effective,glozman1998unified,day2012treatment}
\begin{equation}
v_{\alpha}=V^{\text{(conf)}}_{\alpha}+V^{\text{(hf)}}_{\alpha}~,
\end{equation}
where the confinement potential has the linear form
\begin{equation}
V^{\text{(conf)}}_{\alpha}=V_{0}+C \xi_{\alpha}~.
\end{equation}
The hyperfine potential consists of the pseudo-scalar meson exchanges for the octet
\begin{eqnarray}
V^{\text{(hf)}}_{\alpha}(\mathbf{\xi}_{\alpha}) &=&
\sum_{F=1}^{3} V_{\pi} ({\mathbf  \xi}_{\alpha})  \lambda_{\beta}^{F} \lambda_{\gamma}^{F} \,  
{\bf{\sigma}}_{\beta}  \cdot {\bf \sigma}_{\gamma} 
\nonumber \\
&& + \sum_{F=4}^{7} V_{K} ({\bf  \xi}_{\alpha})  \lambda_{\beta}^{F} \lambda_{\gamma}^{F} \, 
{\bf  \sigma}_{\beta} \cdot {\bf \sigma}_{\gamma} 
\nonumber \\
&& + V_{\eta_8} ({\bf  \xi}_{\alpha})  \lambda_{\beta}^{8} \lambda_{\gamma}^{8}  \,  
 {\bf  \sigma}_{\beta}  \cdot{\bf \sigma}_{\gamma},
 \label{octet}
\end{eqnarray}
and for the singlet
\begin{equation}
V^{\text{(hf)}}_{\alpha}({\bf \xi}_{\alpha}) = 
 \frac{2}{3} V_{\eta_0} ({\bf \xi}_{\alpha}) \;  
 {\bf \sigma}_{\beta} \cdot {\bf  \sigma}_{\gamma},
 \label{singlet}
\end{equation}
 where $\mathbf{  \sigma}$ and $\lambda^{F}$ are the quark spin and flavor matrices, respectively.

So, in an angular momentum basis we get a flavor dependent quark-quark interaction between light quarks
\begin{equation}
V^{\text{u(d)-u(d)}}_{\alpha} = \left\{  p_{\alpha}^{u(d)-u(d)} V_{\pi} T_{\alpha}  +\frac{1}{3}V_{\eta_8}+ \frac{2}{3}V_{\eta_0}\right\} 
\Sigma_{\alpha},
  \label{vuu}
\end{equation}
between light and strange quarks
\begin{equation}
V^{\text{u(d)-s}}_{\alpha} =\left\{ 2 p_{\alpha}^{u(d)-s}  V_{K} -\frac{2}{3}V_{\eta_8}+ \frac{2}{3}V_{\eta_0}  \right\} \Sigma_{\alpha}, 
\label{vus}
\end{equation}
and between strange quarks
\begin{equation}
V^{\text{s-s}}_{\alpha} = \left\{ \frac{4}{3}V_{\eta_8}+ \frac{2}{3}V_{\eta_0}  \right\} \Sigma_{\alpha}.
\label{vss}
\end{equation}
The quarks are spin-$1/2$ particles and the isospin of the light quarks is $1/2$, while the isospin of the strange 
quark equals $0$. Therefore, the symmetry coefficients are
\begin{equation}
p_{\alpha}^{u(d)-u(d)}=(-1)^{l_{\alpha}+s_{\alpha}+\tau_{\alpha}-2}
\end{equation}
and
\begin{equation}
p_{\alpha}^{u(d)-s}=(-1)^{l_{\alpha}+s_{\alpha}-1},
\end{equation}
and the $T_{\alpha}$ and $\Sigma_{\alpha}$ isospin-isospin and spin-spin factors  are given by
\begin{equation}
T_{\alpha} =  2 \tau_{\alpha} (\tau_{\alpha}+1)-3
\end{equation} 
and 
\begin{equation}
\Sigma_{\alpha} =  2 s_{\alpha} (s_{\alpha}+1)-3,
\end{equation}
respectively.

\section{Faddeev approach to three-quark problems}
\label{SEC:Faddeev}
The Faddeev approach is exceptionally well suited for three quark problems. It can take into account the short-range correlations,
the confinement asymptotics and the possible exchange symmetry of particles.  
To accomplish this, we split the quark-quark potential
into confining and non-confining terms 
\begin{equation}
v_{\alpha}= v_{\alpha}^{(c)} + v_{\alpha}^{(s)},
\end{equation}
where $c$ and $s$ stand for confining and short-range, respectively \cite{papp1999treatment,PhysRevC.62.044004,day2012treatment}.
Then the Schr\"odinger equation takes the form
\begin{equation}\label{schrodinger}
H|\Psi\rangle=(H^{(c)}+v_{1}^{(s)}+v_{2}^{(s)}+v_{3}^{(s)} )|\Psi \rangle = E |\Psi\rangle~,
\end{equation} 
with 
\begin{equation}
H^{(c)}= H^{(0)}+v_{1}^{(c)}+v_{2}^{(c)}+v_{3}^{(c)}.
\label{Hc}
\end{equation}

The Faddeev method amounts to splitting the  wave function $|\Psi\rangle$ naturally  into three 
components 
\begin{equation}\label{fdec2}
|\Psi\rangle=|\psi_{1} \rangle + |\psi_{2} \rangle + |\psi_{3} \rangle~.
\end{equation} 
The Faddeev components satisfy the equation
\begin{equation}\label{fcomp}
|\psi_{\alpha}\rangle=G^{(c)}_{\alpha}(E)v_{\alpha}^{(s)} ( |\psi_{\beta} \rangle + |\psi_{\gamma} \rangle) , \ \ \ \ \alpha=1,2,3,
\end{equation}
where the  channel Green's operators are given by
\begin{equation}
G_{\alpha}^{(c)}(E)=(E-H^{(c)}-v_{\alpha}^{(s)})^{-1}.
\end{equation}

We need to introduce the appropriate orbital angular momentum basis. The orbital angular momentum associated with coordinates $\xi_{\alpha}$ and 
$\eta_{\alpha}$ are denoted by $l_{\alpha}$ and $\lambda_{\alpha}$, respectively, and they are coupled to the total orbital angular momentum $L$. The spin of
particles $\beta$ and $\gamma$, $S_{\beta}$ and $S_{\gamma}$, respectively, are coupled to $s_{\alpha}$, which is with 
the spin of particle $\alpha$, $S_{\alpha}$, 
coupled to the total spin $S$. Similarly, the isospin of
particles $\beta$ and $\gamma$, $t_{\beta}$ and $t_{\gamma}$, respectively, are coupled to $\tau_{\alpha}$, which is with the isospin of particle 
$\alpha$, $t_{\alpha}$, 
coupled to the total isospin $T$. The angular momentum $L$ and spin $S$ are coupled to total angular momentum $J$. So, 
we adopted $LS$ coupling, which is appropriate if the quark-quark interaction does not have tensor terms.

A further advantage of the Faddeev method is that the identity of particles greatly simplifies the equations.
If particles $\beta$ and $\gamma$ are identical, the wave function $\Psi$ must be symmetric with respect to the exchange of these particles. 
We denote ${\cal P}_{\alpha}$ as the operator that exchanges particles $\beta$ and $\gamma$. Then 
\begin{equation}
{\cal P}_{\alpha}|\Psi \rangle = p_{\alpha}|\Psi\rangle
\end{equation}
where 
\begin{equation}
p_{\alpha}=(-1)^{l_{\alpha}+s_{\alpha}-S_{\beta}-S_{\gamma} + \tau_{\alpha} - t_{\beta} -t_{\gamma}}~,
\label{sym}
\end{equation}
if particles carry spin and isospin.

Putting everything together and assuming that particles $2$ and $3$  
are identical, the three-component Faddeev equations simplify to
\begin{equation}
\begin{pmatrix}
|\psi_{1} \rangle \\
|\psi_{2} \rangle
\end{pmatrix} =
\begin{pmatrix}
0 & 2  G_{1}^{(c)} v_{1}^{(s)}  \\
G_{2}^{(c)}  v_{2}^{(s)} &  G_{2}^{(c)} v_{2}^{(s)} p_{1} {\cal P}_{1}
\end{pmatrix}
\begin{pmatrix}
|\psi_{1} \rangle \\
|\psi_{2} \rangle
\end{pmatrix}.
\label{2identicalfe}
\end{equation}

If all three particles are identical, Eq.~(\ref{2identicalfe}) gets further reduced to one single equation
\begin{equation} \label{fmp}
| \psi_{1} \rangle =  2 G_1^{(c)} v_1^{(s)} {\mathcal P}_{123}
| \psi_{1} \rangle,
\end{equation}
where ${\mathcal P}_{123}={\mathcal P}_{12}{\mathcal P}_{23}$ is the operator for cyclic permutation of all three particles 
${\mathcal P}_{123} |\psi_{1}\rangle =| \psi_{2}\rangle$. 
\\

\section{Results for light and strange baryons}
\label{SEC:LightAndStrangeBaryons}
In  the GBE model, the spatial part of the potential is given as a sum of two Yukawa potentials
\begin{equation}
V_{\gamma} = \frac{g^{2}_{\gamma}}{4 \pi} \frac{1}{12 m_{i} m_{j}} 
 \left[ \mu^{2}_{\gamma}  \frac{\exp( -\mu_{\gamma} r_{i j})}{ r_{i j}}  - 
  \Lambda^{2}_{\gamma}  \frac{\exp( -\Lambda_{\gamma} r_{ij})}{r_{ij} }\right],
\label{potential}
\end{equation}
where $\gamma$ stands for the exchange mesons  $\pi$, $K$, $\eta$ and $\eta'$. The $\Lambda_{\gamma}$ term is assumed to have 
a linear dependence on meson masses
\begin{equation}
\Lambda_{\gamma} = \Lambda_{0}+ \kappa \mu_{\gamma},
\label{eq:lg}
\end{equation}
with $\Lambda_{0}$ and $\kappa$ fitted parameters.  In this model the masses of the constituent quarks and exchange mesons 
are fixed parameters in the vacuum calculation, $m_{u}=m_{d}=340$ MeV, $m_{s} = 500$ MeV, $\mu_{\pi}=139$ MeV, $\mu_{K}=494$ MeV, $\mu_{\eta}= 547$ MeV, 
$\mu_{\eta'} = 958$ MeV, and the meson octet-quark coupling constant $g^{2}_{8}/4\pi = 0.67$.
The other vacuum parameters were obtained by fitting the model to the experimental baryon spectrum and excellent agreement was found with 
$V_{0} = -416$ MeV, $C=2.33\ \mbox{fm}^{-2}$, $\Lambda_{0} = 2.86\ \mbox{fm}^{-1}$, $\kappa = 0.81$ and $(g_{0}/g_{8})^{2}=1.34$.

If $H^{(c)}$ in Eq.\ \eqref{Hc} has bound states,
we may get non-vanishing Faddeev components from a vanishing total wave function. 
These are spurious states. To avoid them, we should push the possible bound states of $H^{(c)}$ out of the spectrum of physical interest.
We take 
\begin{equation}
v^{(c)} = V^{(conf)} + a_{0}\exp(-(r/r_{0})^{2})
\end{equation}
and
\begin{equation}
v^{(s)} = V^{(hf)} - a_{0}\exp(-(r/r_{0})^{2}),
\end{equation}
with $a_{0}= 3\ \mbox{fm}^{-1}$ and $r_{0}= 1\ \mbox{fm}$. With this choice of parameters we guarantee the absence of spurious 
solutions in the energy range below $\approx 2$ GeV.

\section{In-Medium Scaling of Model Parameters}\label{sec:scaling}

This study investigates how strongly the constituent quark model responds to parameter changes that could arise 
from a medium dependence, i.e. due to finite temperature and/or density effects. 
In the GBE model, a single interaction potential naturally handles the different masses of the light and strange quarks.
This mass dependence is a promising entry point for exploring medium effects that vary the constituent quark masses. 
To narrow our scope, we make a series of simplifying assumptions:
\begin{enumerate}[wide, labelindent=0pt]
    \item We assume that all model parameters vary as a function of temperature $T$ and chemical potential $\mu$ only through their dependence on the light quark mass gap $\Sigma^*$
    \begin{equation}
        m^*_q = m_q^0 +\Sigma^*,
    \end{equation}
    where $\Sigma^*$ tends from its vacuum value $\Sigma = m_q - m_q^0$ to zero.
    In analogy to a Nambu--Jona-Lasino (NJL) model \cite{nambu1961dynamical1, nambu1961dynamical2, Klevansky:1992qe}, we associate the mass gap directly with the light quark mean-field condensate $\Sigma^* \propto \langle \bar{q} q \rangle ^*$.

    \item We take the strange quarks to vary as 
    \begin{equation}
       m^*_s = m_s^0 +\alpha \Sigma^*,
    \end{equation}
    where $\alpha \approx 1.2$ is set to match vacuum constituent mass and the current mass of the strange quark.
    This is a deliberate oversimplification that,  
    while masking underlying dynamics, 
    makes simple scaling relations tractable.

    \item  We hold constant as a function of $\Sigma^*$ several model parameters: 
    $g_0/g_8$, the ratio between the singlet and octet couplings;
    $\Lambda_0$ and $\kappa$, which control the meson-mass dependent smearing of the delta function;
    and $V_0$, the offset energy.
    Section \ref{sec:results} discusses the effect of lifting these assumptions.
\end{enumerate}

The meson-quark-quark couplings $g_\gamma^*$, 
exchange boson masses $\mu_\gamma^*$, 
and the linear coefficient of confinement $C^*$ are taken to scale in medium as power laws of $\Sigma^*$:
\begin{align}
g_\gamma &\rightarrow g_\gamma^* \propto (\Sigma^*)^{\nu_g} \\
\mu_\gamma &\rightarrow \mu_\gamma^* \propto (\Sigma^*)^{\nu_\mu} \\
C &\rightarrow C^* \propto (\Sigma^*)^{\nu_c}
\end{align}
We calculate the baryon spectra as a function of $\Sigma^*$ for different choices of scaling exponents $\nu_g$, $\nu_\mu$, and $\nu_c$. 

We organize our study into different `schemes' which determine $\nu_g$ and $\nu_\mu$, while $\nu_c$ is handled separately.
See Table \ref{tab:schemes}.
For each scaling scheme, we consider $\nu_c = 1$ and $\nu_c = 0$.
The former is motivated by the expectation that confinement weakens with chiral restoration.
The latter neglects the weakening of confinement.

\begin{table}
\centering
\caption{Scaling schemes used in this study. The table shows the scaling of the coupling constants 
$g_{\gamma}^* \propto (\Sigma^*)^{\nu_g}$ and meson masses $\mu_\gamma^* \propto (\Sigma^*)^{\nu_\mu}$ for each scheme. 
Schemes C-F are determined by the scaling of the pion decay constant $f_\pi^* \propto (\Sigma^*)^{a}$. 
Schemes O-B do not.
}
\setlength{\tabcolsep}{6pt}
\begin{tabular}{l|*{3}{c}*{4}{c}}
\hline
& O & A & B & C & D & E & F \\
\hline
$\nu_{g}$ & $0$ & $3/2$ & $1$ & $1$ & $2/3$ & $1/2$ & $0$ \\
$\nu_\mu$ & $0$ & $0$ & $1$ & $1/2$ & $1/6$ & $0$ & $-1/2$ \\
$a$ & -- & -- & -- & $0$ & $1/3$ & $1/2$ & $1$ \\
\hline
\end{tabular}
\label{tab:schemes}
\end{table}

\begin{figure*}[t]
\centering
\includegraphics[width=1.\linewidth]{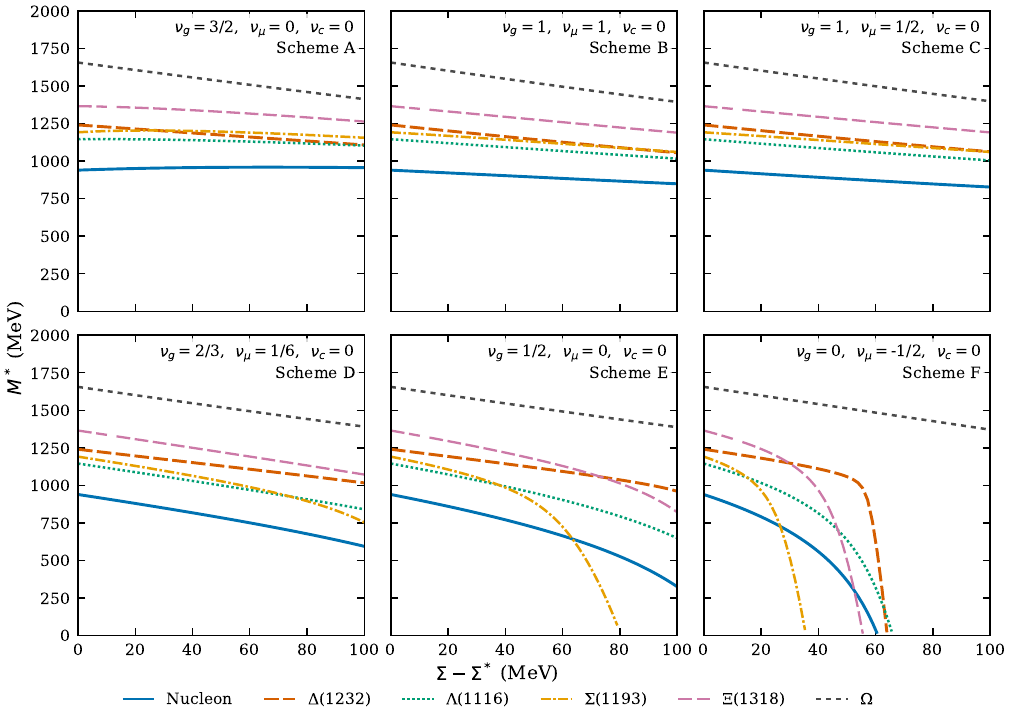}
\caption{
Comparison of Schemes A-F. Baryon spectra as a function of $\Sigma-\Sigma^*$, with the
linear confinement term not scaled ($\nu_c = 0$).
The lowest lying state of each baryon family is plotted.}
\label{fig:scan0}
\end{figure*}

\begin{figure*}[t]
\centering
\includegraphics[width=1.\linewidth]{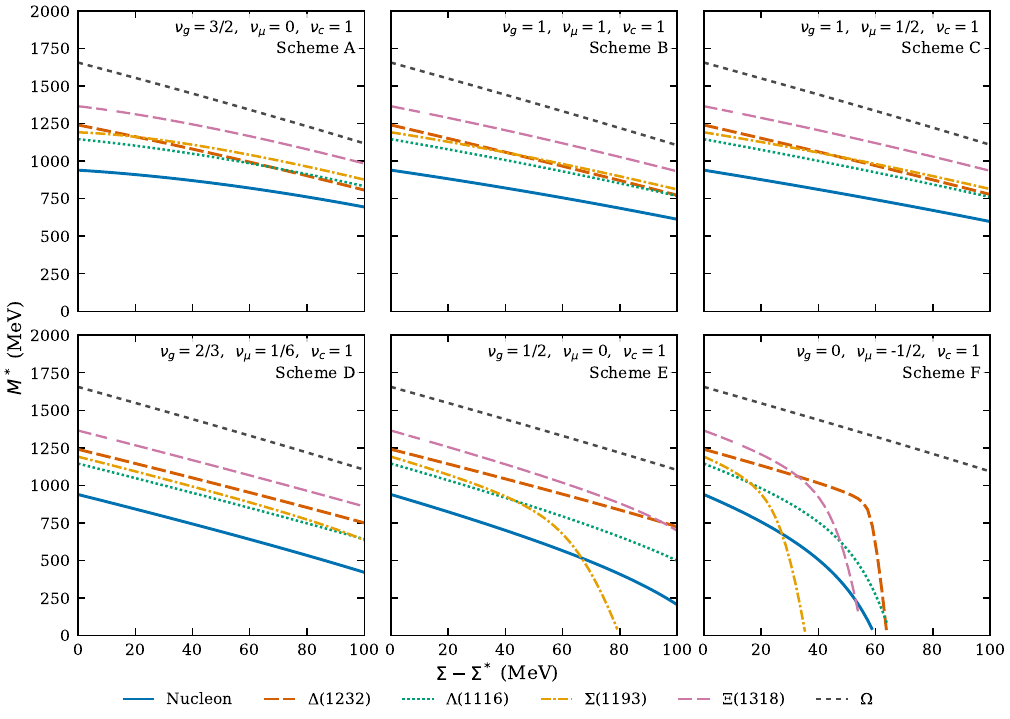}
\caption{
Comparison of Schemes A-F. Baryon spectra as a function of $\Sigma-\Sigma^*$, with the
linear confinement term scaled linearly ($\nu_c = 1$).
The lowest lying state of each baryon family is plotted.
}
\label{fig:scan1}
\end{figure*}

\begin{figure}
\centering
\includegraphics[width=0.95\linewidth]{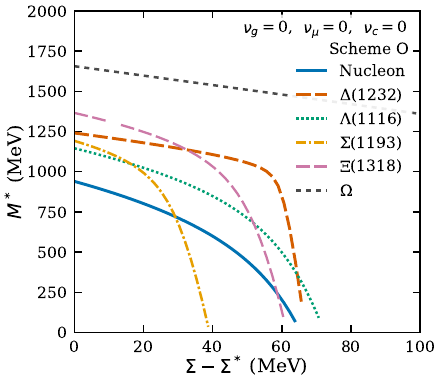}
\caption{
Baryon spectra as a function of $\Sigma-\Sigma^*$ for baseline scheme O, which keeps all model parameters constant besides the constituent quark masses.
The lowest lying state of each baryon family is plotted.}
\label{fig:scho}
\end{figure}

\begin{figure*}[t]
\centering
\includegraphics[width=0.95\linewidth]{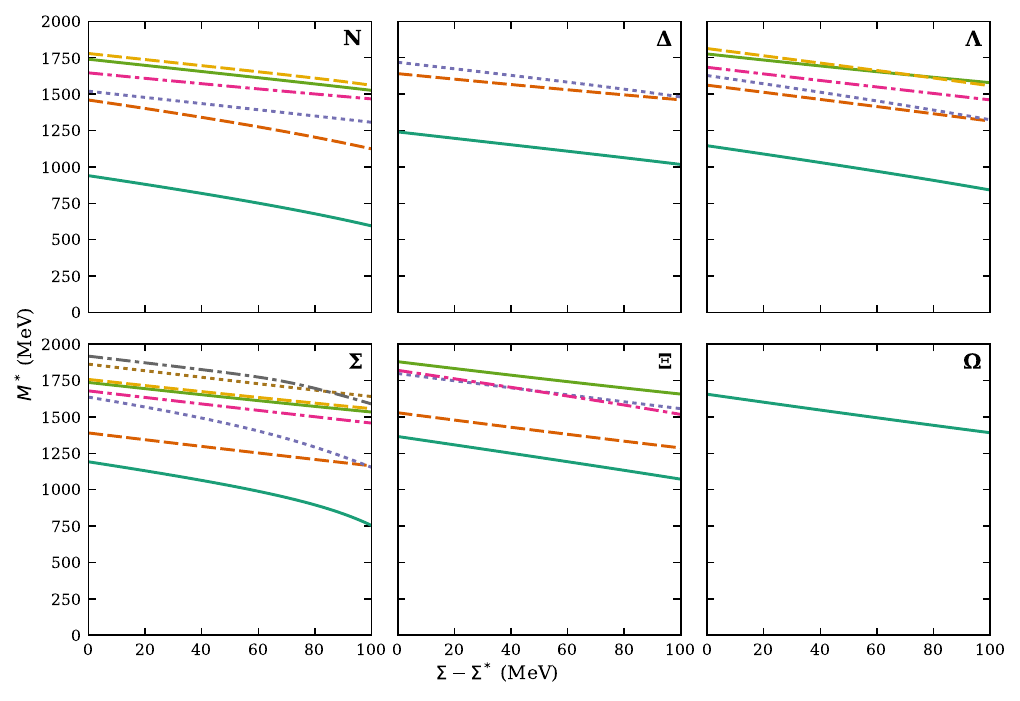}
\caption{
All calculated baryons masses shown as a function of $\Sigma-\Sigma^*$ for Scheme D ($\nu_g=2/3$, $\nu_\mu=1/6$, $\nu_c=0$).
Plotted baryons listed by family in ascending order of calculated vacuum mass: 
$\boldsymbol{N}$: Nucleon, $N(1440)$, $N(1535)-N(1520)$, $N(1650)-N(1700)-N(1675)$, $N(1680)-N(1720)$, $N(1710)$; 
$\boldsymbol{\Delta}$: $\Delta(1232)$, $\Delta(1620)-\Delta(1700)$, $\Delta(1600)$; 
$\boldsymbol{\Lambda}$: $\Lambda(1116)$, $\Lambda(1405)-\Lambda(1520)$, $\Lambda(1600)$, $\Lambda(1670)-\Lambda(1690)$, $\Lambda(1800)-\Lambda(1830)$, $\Lambda(1810)$; 
$\boldsymbol{\Sigma}$: $\Sigma(1193)$, $\Sigma(1385)$, $\Sigma(1660)$, $\Sigma[1560]-\Sigma(1670)$, $\Sigma[1620]-\Sigma[1775]-\Sigma[1940]$, $\Sigma(1750)$, $\Sigma[1690]$, $\Sigma(1880)$; 
$\boldsymbol{\Xi}$: $\Xi(1318)$, $\Xi(1530)$, $\Xi(1820)$, $\Xi[1690]$, $\Xi[1950]$; 
$\boldsymbol{\Omega}$: $\Omega$.
States that are degenerate in the model are listed together.
}
\label{fig:schD}
\end{figure*}

\begin{figure}[t]
\centering
\includegraphics[width=0.9\linewidth]{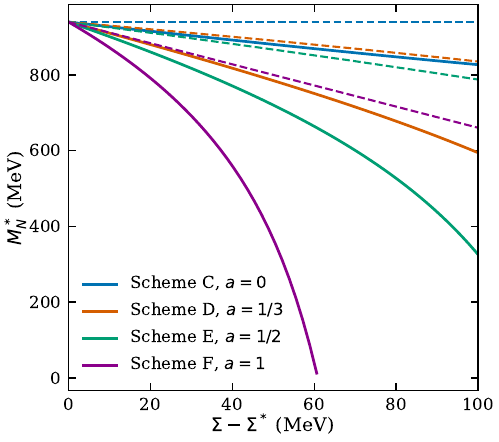}
\caption{
Comparison of the calculated nucleon mass $M_{N,calc}^*$ (solid lines) and the nucleon mass obtained from scaling relations, Eqs.~\ref{BR1} and \ref{BR2}, given by
$M_{N,scale}^* =  M_N \times (\Sigma^*/\Sigma)^a$ (dashed lines).
Results are shown for schemes C-F, with $\nu_c=0$.
}
\label{fig:BR}
\end{figure}

The first three scaling schemes are chosen to provide a baseline and to isolate the effects of varying $g_\gamma^*$ and $\mu_\gamma^*$ from each other.
Scheme O does not scale the couplings or meson masses.
Scheme A approximately scales the interaction potential linearly $V_{\gamma} \propto \Sigma^*$.
Scheme B scales both the coupling constants and meson masses linearly.

In schemes C-F, we constrain the couplings $g_\gamma^*$ and exchange boson masses $\mu_\gamma^*$ to vary based on a single dependence on the behavior of the pion decay constant $f_\pi^*$.
The scaling relations in these schemes are inspired by Brown-Rho scaling \cite{Brown1991, Brown1996, brown2002manifestation}.
We work with the approximations that the Gell-Mann-Oakes-Renner relation \cite{GellMann1968}
\begin{equation}
(\mu_{\pi}^*)^2 (f_\pi^*)^2 = -2 m_q^0 \langle \bar{q} q \rangle^* + O\left((m_q^0)^2\right)
\label{eqn:gmor}
\end{equation}
and the constituent quark-level Goldberger-Treiman relation
\begin{equation}
g_{\pi qq}^* f_\pi^* = m_q^* g_A^Q
\label{eqn:gt}
\end{equation}
hold for in-medium values far from chiral restoration,
and take the axial charge of the constituent quark to be $g_A^Q = 1.0$.
These approximations hold when working at the tree level and in a mean field approximation, and are consistent, for example, with a two-flavor NJL model \cite{hatsuda1994qcd}.

We make the Ansatz that the pion decay constant varies with the quark condensate in a power law relationship
\begin{align}
f_\pi^* &\propto \left(\langle \bar{q} q \rangle ^*\right)^a.
\label{eq:f_ans}
\end{align}
In the different parameterization schemes C-F we explore different Ans\"atze for the value of the exponent $a$. 
Approximating $m^*_q \propto \langle \bar{q} q \rangle ^*$ sets the pion-quark-quark coupling 
and pion mass
to vary with the quark condensate as
\begin{align}
g_{\pi qq}^* & \propto\left(\langle \bar{q} q \rangle^*\right)^{1-a} \\ 
\mu_\pi^* & \propto \left(\langle \bar{q} q \rangle^*\right)^{1/2-a}.
\end{align}

Finally, for simplicity and tractability we make the Ansatz that these scaling relations can be extended to each coupling constant and meson mass equivalently: 
\begin{equation}
\begin{aligned}
g_\gamma^* & \propto (\Sigma^*)^{1-a} \\ 
\mu_\gamma^* & \propto (\Sigma^*)^{1/2-a}
\end{aligned}
\label{eq:scaling2}
\end{equation}
An alternative choice of extending the scaling only to the octet mesons and not the singlet $\eta'$ will be discussed in Section \ref{sec:results}.
Schemes C-F explore $a=0$, $1/3$, $1/2$, and $1$.
 
The GBE constituent quark model should not be expected to hold at large deviations from vacuum. 
At higher energy scales, gluon degrees of freedom become increasingly important at 
mediating interactions between quarks,
which is neglected in the GBE potential.
Additionally, the scaling relations were motivated by current algebra relations (Eqs. \eqref{eqn:gmor} and \eqref{eqn:gt})
which neglect contributions of order $(\mu_\pi^*/2m_q^*)^2$ and $m_q^0/m_q^*$ \cite{hatsuda1994qcd}. 
As such, we restrict our calculation to the regime $\Sigma-\Sigma^* < 100$ MeV, corresponding to a constituent light quark mass $m_q^* > 240$ MeV.

\section{Results and Discussion} \label{sec:results}

For Schemes A-F,
the lowest lying states in each family are plotted in Fig.~\ref{fig:scan0} for $\nu_c=0$ and 
Fig.~\ref{fig:scan1} for $\nu_c=1$.
Fig.~\ref{fig:scho} shows the behavior when only the constituent quark masses are scaled.
For Scheme D, 
all calculated states from the $N$, $\Delta$, $\Lambda$, $\Sigma$, $\Xi$, and $\Omega$ families are plotted in Fig.~\ref{fig:schD}.
We observe a decreasing baryon mass as constituent quark mass decreases for nearly all baryons across considered scaling schemes. 
The dominant effect on the baryon spectra comes from the scaling of 
the couplings
$g_\gamma^*$. The scaling of the confinement parameter $C^*$ has a smaller but noticeable effect, and the scaling of the meson masses $\mu_\gamma^*$ has minimal effect.

The effect of scaling $g_\gamma^*$ can be understood by inspecting the leading prefactor of the potential (Eq. \eqref{potential}),  
$V_\gamma \propto (g_\gamma^*/m_q^*)^2$. 
When $g_\gamma^*$ scales faster than $m_q^*$, as occurs in scheme A, the total strength of the hyperfine interaction 
decreases as chiral symmetry is restored. However, if $g_\gamma^*$ scales slower than $m_q^*$, the opposite occurs,
as is the case in schemes D-F.

For schemes E and F, we observe a breakdown of the model occurring at relatively small deviations from vacuum,
marked by several baryon states reaching negative energies, a nonphysical result. 

To investigate whether this behavior is caused by the assumptions made in Section \ref{sec:scaling}, 
we test a few simple alternate assumptions for the in-medium scaling of the offset energy $V_0$, 
delta function smearing $\Lambda_\gamma$, 
singlet meson mass $\mu_{\eta'}$,
and singlet coupling $g_0$.
Table \ref{tab:alt_scaling} shows the effect these alternative assumptions have upon the pathology.
While these alternatives do affect the pathology,
none removes it,
indicating this behavior is not solely an artifact of these choices.

\begin{table}[h]
\centering
\caption{Alternative scaling choices 
and their effect on onset of model breakdown,
marked by the last point where the nucleon mass $M_N^*$ is positive.
The five studied cases: 
1. Scaling $V^{conf,*} \propto (\Sigma^*)^{\nu_c}$, achieved by  scaling the parameter $V_0^* \propto (\Sigma^*)^{\nu_c}$.
2. Scaling $\Lambda_\gamma^* \propto (\Sigma^*)^{\nu_\mu}$, achieved by scaling the parameter $\Lambda_0^* \propto (\Sigma^*)^{\nu_\mu}$.
3. Holding $\Lambda_\gamma^*$ constant, achieved by scaling the parameter $\kappa^* \propto (\Sigma^*)^{-\nu_\mu}$.
4. Holding $\mu_{\eta'}^*$ constant, instead of scaling it with the octet meson masses.
5. Holding $g_0^*$ constant, instead of scaling it with the octet couplings.
In each case the modification is applied to scheme F ($\nu_c = 1$), 
except for the final case, in which it is applied to scheme E ($\nu_c = 1$).
}
\label{tab:alt_scaling}
\begin{ruledtabular}
\begin{tabular}{llll}
Case & Onset of breakdown, $\Sigma-\Sigma^*$ \\
\hline

$V^{conf,*}$ is scaled
& $59 \rightarrow 64~\mathrm{MeV}$ \\

$\Lambda_\gamma^*$ is scaled
& $59 \rightarrow 45~\mathrm{MeV}$ \\

$\Lambda_\gamma^*$ held constant
& $59 \rightarrow 50~\mathrm{MeV}$ \\

$\mu_{\eta'}^*$ held constant
& $59 \rightarrow 59~\mathrm{MeV}$ \\

$g_0^*$ held constant 
& $113 \rightarrow 131~\mathrm{MeV}$ \\

\end{tabular}
\end{ruledtabular}
\end{table}

Two other possible causes of the pathology are the initial assumption that the constituent quark model 
can be extended to medium in this manner and the choice of scaling scheme. 
The minimal effect that meson mass scaling has on the baryon spectra implies the pathology is a result of the relative scaling of $g_\gamma^*$ and $m_q^*$,
and in particular does not depend on the current algebra relations motivating schemes C-F.
Thus, our result can be considered as evidence against certain scaling relations for $g_{\pi qq}^*$ holding at small deviations from vacuum,
as a bound on the region of validity of an ad hoc extension of the constituent quark model to medium, 
or both.

An interesting point of comparison is to Brown-Rho scaling, which proposes that the nucleon mass varies linearly with the pion decay constant
\begin{align}
\frac{f^*_\pi}{f_\pi} \approx \frac{M_N^*}{M_N}.
\label{BR1}
\end{align}
Different scaling relations between the pion decay constant and the quark condensate 
\begin{align}
\frac{f^*_\pi}{f_\pi} 
&\approx 
\left(\frac{\langle \bar{q} q \rangle^*}{\langle \bar{q} q \rangle}\right)^a 
\label{BR2}
\end{align}
have appeared in the Brown-Rho literature. Early work suggested the scaling exponent to be $a=1/3$ \cite{Brown1991}. Later, scaling with $a=1/2$ \cite{Brown1996}, which maintains a constant pion mass, and $a=1$ \cite{koch1993model},  Nambu scaling, were explored.
In schemes C-F, we make a matching Ansatz (Eq. \eqref{eq:f_ans}) for the scaling of the pion decay constant. 
In our calculation this Ansatz fixes the scaling relations for $g_\gamma^*$ and $\mu_\gamma^*$, with which we calculate $M_N^*$. 
This gives us a natural point of comparison between the in-medium nucleon mass that we calculate and the mass predicted by Brown-Rho scaling.
This comparison is shown in Fig.~\ref{fig:BR}, 
where we have also included the $a=0$ case as a baseline. 
We observe the nucleon mass decreasing faster for larger values of the scaling exponent $a$, as would be expected.
However, in each scheme the calculated change in the nucleon mass is greater than these scaling relations would predict
\begin{align}
\left( \frac{M_N^*}{M_N}\right)_{calc} \lesssim  \left(\frac{\Sigma^*}{\Sigma}\right)^a.
\end{align}

\section{Sensitivity of ideal-gas estimates to baryon mass defects}
\label{SEC:Thermo}

The main focus of this work is the study of medium-motivated changes of the hadron spectrum as outlined in the previous sections.
In this last section we illustrate how sensitive particle yields and yield ratios can be to baryon mass shifts of the size generated in the model study above.
In this context, it is important to us to emphasize that we add this section for illustration of possible effects without 
the intent to offer a full, self-consistent thermodynamic model.

As a preliminary exploration of the thermodynamic implications of this calculation, we note the elementary consequence of a mass shift on an ideal gas.
The HRG model, in its traditional form, assumes a non-interacting relativistic gas of hadrons, 
with each species having constant mass $M$ at all temperatures and chemical potentials. 
We take this as a guide, and make the trial assumption that at some temperature this system can still be described as an ideal gas, 
but with a shifted effective mass $M^*$.

For baryons, their relatively large mass motivates the taking of non-relativistic $p \ll M$ and Boltzmann $n \lambda_{TH}^3 \ll 1$ limits. 
At zero chemical potential ($\mu=0$) for a single species ideal gas, 
the ratio of shifted to unshifted pressure and of shifted to unshifted  density is, to first order:
\begin{equation}
\frac{P_{M^*}}{P_{M}} \approx 
 \left(\frac{M^*}{M}\right)^{3/2}\exp\left(\frac{M-M^*}{T}\right).
\end{equation}

In order to quantify the effect of
dynamically reduced baryon masses, we introduce the dimensionless quantity
\begin{equation}
    \eta=\frac{M-M^*}{T}.
\end{equation}
For small values of $\eta$
an expansion results in 
\begin{equation}
    \ln\frac{P_{M^*}}{P_M}
    =
    \eta+\frac{3}{2}
    \ln\left(
    1-\eta\frac{T}{M}
    \right)
    \approx \eta
    \left(
    1-\frac{3T}{2M}
    \right).
\end{equation}
Since $\eta$ is positive, for masses 
\begin{equation}
    M>\frac{3}{2}T
\end{equation}
we expect an increase of the ideal-gas pressure.
At a crossover temperature of $T_c=155 \;\rm MeV$, this condition is satisfied for all known baryons.
Further, this enables us to estimate what value of $\eta$ would result in negligible changes $\varepsilon$ of the pressure, viz.
\begin{equation}
    \left|
    \frac{P_{M^*}}{P_M}-1
    \right|
    \lesssim\varepsilon.
\end{equation}
We obtain approximately
\begin{equation}
    \eta\lesssim\frac{\ln(1+\varepsilon)}{1-\frac{3T}{2M}}.
\end{equation}
For the neutron at $T=155\;\rm MeV$, we find that the ideal-gas pressure changes less than $(5,10,20)\%$ if the difference between vacuum and in-medium mass is less than $(10,20,37)\;\rm MeV$. 
The changes in pressure at fixed $\eta$ are more pronounced for heavier baryons.
In general, the effect in the baryonic sector seems negligible only for $\eta\ll 0.1$, which near the expected crossover temperature $T_c\approx 155$~MeV corresponds to mass-shifts of the order $10-20\;\rm MeV$ or less.

This rough approximation neglects any self-consistent thermodynamic picture and any interactions between hadrons. 
In most relevant cases, this shifted mass ideal gas without interactions will have a higher pressure than an ideal HRG. 
It is however known from RMF models, that interaction contributions to the pressure, for instance those of exchange-mesons, tend to counter the pressure increase due to chiral symmetry restoration.
The full impact of medium-dependent quark masses on the thermodynamics of the full baryon ensemble therefore cannot be determined without specifying the underlying interaction contributions to the net pressure. 

If mass shifts of only a few $10$~MeV were inserted into an ideal HRG description, they would be amplified by exponential Boltzmann factors and produce non-negligible changes in hadron yields.
We illustrate this by approximating the number densities as an ideal Boltzmann gas with
\begin{equation}
\label{EQ:Yield}
    n_i=g_i\left(\frac{M_iT}{2\pi}\right)^{3/2}
    e^{-\frac{M_i-\mu_i}{T}}.
\end{equation}
We expect non-negligible modifications of the nucleon yield, even at moderate differences between vacuum and in-medium masses as these differences enter exponentially, viz.
\begin{eqnarray}
\label{EQ:relativeYield}
    \frac{n_i^*}{n_i}=\left(\frac{M_i^*}{M_i}\right)^{3/2}
     e^{\frac{M_i-M_i^*}{T}}.
\end{eqnarray}
The yield ratios between different baryons change accordingly as
\begin{equation}
\frac{
    \left(
    n_i/n_j
    \right)^*
    }{
    \left(
    n_i/n_j
    \right)    
    }
    =
    \left(\frac{M_i^*/M_i}{M_j^*/M_j}\right)^{3/2}
    e^{\frac{\Delta M_i-\Delta M_j}{T}}.
\end{equation}
If $\Delta M_i=M_i-M_i^*$ is of the same magnitude as $\Delta M_j$ this relation implies only small variations of the yield ratios due to cancellations in the exponent which leave the prefactor as the dominant source of variations.

\begin{figure}
\centering
\includegraphics[width=0.95\linewidth]{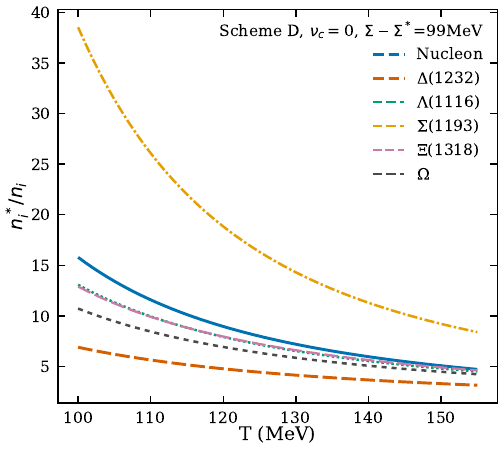}
\caption{
\label{FIG:YieldDT}
Effect of moderate changes in the baryon mass spectrum (illustrated with scheme~D, $\Sigma-\Sigma^*=99$ MeV):
Relative baryon yields $n(M^*_i,T)/n(M_i,T)$ at constant $\Sigma-\Sigma^*$ 
at varying temperature. 
Lower temperatures show larger yield deviations.
}
\end{figure}

We illustrate yield variations as described by Eq.(\ref{EQ:relativeYield}) with scheme D that shows only a moderate constituent quark mass dependence. Here, the $\Sigma$-baryon with the steepest mass decrease shows the strongest deviations from particle yields as calculated with constant $\Sigma$-mass.
As Fig.~\ref{FIG:YieldDT} shows, deviations are more pronounced with decreasing temperature. 
Practically, this can result in a competition of a temperature induced yield enhancement with yield stabilization due to reduced baryon mass defect. As we do not offer an explicit relation for the temperature dependence of the constituent quark masses, further elaboration on this effect would go beyond the setup of this work.

For yield ratios, the cancellation of mass defects of two different baryons in the exponential of Eq.~(\ref{EQ:relativeYield}) can lead to minimal deviations from the yield ratios in a constant mass model. Fig.~\ref{FIG:YieldRatioD} illustrates this with Scheme D. At the same time, the figure shows by example of $\Sigma$ and $\Delta$, that mass defects are not guaranteed to cancel and thus can lead to notable deviations from constant mass yield ratios. In general, however, yield ratios show smaller deviations than the particle yields themselves.

As a final note, we illustrate how yields and yield ratios can differ if mass defects are strongly species-dependent. To do so, we show both quantities for the $\Sigma$ in Scheme F, which develops a steep drop of the baryon mass at $\Sigma-\Sigma^*\approx30$~MeV. We should not interpret this parameter region as a quantitative prediction for the physical $\Sigma$ yield, since the rapid mass drop signals the breakdown of the constituent-quark model.
Rather, we perform a stress test of the ideal-gas estimate; if one baryonic species develops a larger mass defect than the rest of the spectrum, the cancellation in yield ratios no longer occurs.

\begin{figure}
\centering
\includegraphics[width=1.\linewidth]{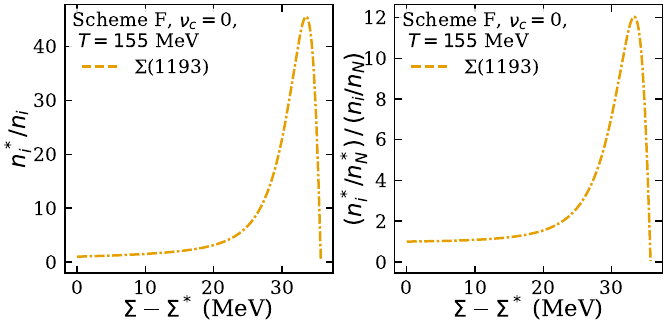}
\caption{
\label{FIG:YieldF}
Effect of significant changes in the baryon mass spectrum (illustrated for $\Sigma$, scheme~F, $T=155$ MeV): 
Left panel: $\Sigma$-yield normalized to constant (vacuum mass) $\Sigma$-yield. 
Larger mass defects result in exponentially increasing yield deviations (see Eq.~(\ref{EQ:Yield})).
At very low baryon masses the mass dependence of the prefactor in Eq.~(\ref{EQ:Yield}) leads to a drop of yield deviations.
\\
Right panel: While yield ratios are not as affected by mass defects due to cancellations, the yield ratio of particles with different mass defects can differ significantly from vacuum mass ideal-gas yield ratios.
While our model is not reliable beyond the peak in both panels, the figure illustrates that strongly species-dependent mass defects can produce notable yield and yield ratio deviations from the ideal-gas behavior with constant baryon masses.
}
\end{figure}

As seen from Fig.~\ref{FIG:YieldF}, both the yield and the yield ratio can deviate by an order of magnitude from the constant-mass expectation. 
At small and moderate mass defects the exponential factor dominates, while at very low baryon masses the explicit mass-dependent prefactor in Eq.~(\ref{EQ:Yield}) and Eq.~(\ref{EQ:relativeYield}) suppresses the yield and yield-ratio estimates. 
While explicitly not intended as a predictive study, we use this extreme and model breaking scenario to demonstrate that the stability of yield ratios, even in domains with moderate mass shifts, is not generic, but crucially relies on similar mass defects among the compared baryons.

\begin{figure}
\centering
\includegraphics[width=0.95\linewidth]{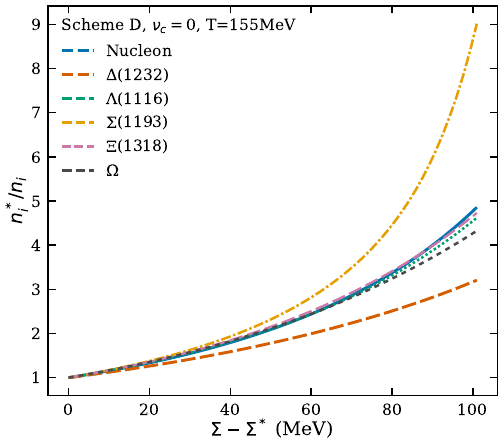}
\caption{
\label{FIG:YieldD}
Effect of moderate changes in the baryon mass spectrum (illustrated with scheme~D, $T=155$ MeV):
Relative baryon yields $n(M^*_i,T)/n(M_i,T)$ as approximated by Eq.~(\ref{EQ:Yield}) for different species (labeled in legend) as an estimate of the effect of in--medium vs. vacuum baryon mass. Notably, small variations of $\Sigma-\Sigma^*$ lead to significant deviations from the vacuum mass yield.
While most baryons show the same yield variation, $\Delta$ and $\Sigma$ notably deviate.
}
\end{figure}

\begin{figure}
\centering
\includegraphics[width=0.95\linewidth]{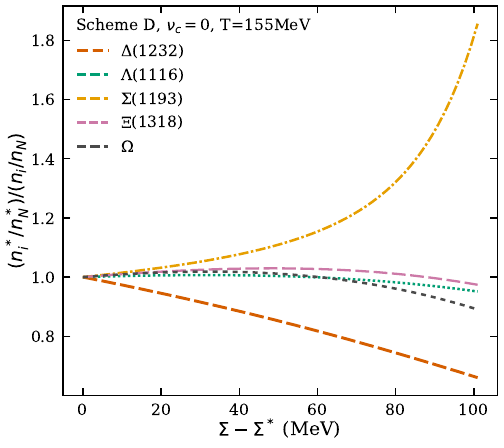}
\caption{
\label{FIG:YieldRatioD}
Effect of moderate changes in the baryon mass spectrum (illustrated with scheme~D, $T=155$ MeV):
The baryon yield relative to nucleon yield, viz. $n(M^*_i)/n(M^*_N)$, normalized to the same relative yield for medium independent (vacuum) mass, viz. $n(M_i)/n(M_N)$. Although the yields from Fig.~\ref{FIG:YieldD} differ from the vacuum mass result, yield deviations similar to the nucleon yield result in yield ratios close to one. As in Fig.~\ref{FIG:YieldD}, this does not hold for $\Delta$ and $\Sigma$. In general, yield ratios are less affected by medium shifted baryon masses than particle yields themselves.
}
\end{figure}

\section{Summary}
\label{SEC:SUMMARY}
We have studied the sensitivity of the light and strange baryon spectrum to medium-motivated changes of the parameters in a GBE relativistic constituent quark model. 
The three-quark bound-state problem was solved in a Faddeev approach, using a vacuum parameterization that reproduces the light and strange baryon spectrum reasonably well below about 2 GeV.
To approximate the effect of chiral symmetry restoration in the quark sector, we accounted for possible scalings of the quark-meson coupling, exchange meson masses, and confinement strength.

Across most scaling schemes, the baryon masses decrease with decreasing constituent quark mass. 
The strongest sensitivity comes from the scaling of the quark-meson coupling.
Changes in the confinement strength have a visible, but smaller effect.
Scaling the exchange meson masses has comparatively little impact.

We observe that some scaling choices drive the spectrum into non-physical negative-energy states. 
This could indicate either a domain of non-applicability of the model itself or of the chosen scaling relations.
The calculated nucleon masses generally drop faster than suggested by Brown-Rho-type scaling.
It seems worthwhile to investigate whether a more sophisticated treatment,
such as proper handling of strange quark condensate,
different in-medium handling of the delta function in the GBE potential, 
or alternative scaling choices, 
could remove the pathological behavior or match Brown-Rho scaling.  
As an illustration of possible consequences, we estimated the effect of changes in the mass spectrum on particle yields 
and compared them to ideal-gas baryon yields.
Even mass shifts of a few tens of MeV change absolute yields notably. 
The effect on yield ratios is less pronounced unless the shifts are strongly species-dependent.
Overall, this work should be understood as a sensitivity study which shows that medium-dependent 
baryon-spectra can respond strongly to implied scaling laws. 
A self-consistent connection to thermodynamics and a full in-medium hadron-resonance-gas treatment remain natural next steps.

\section*{Acknowledgments}

We are thankful to Dr.~C.~D.~Roberts for discussions and helpful comments. 

\bibliography{3q}
\bibliographystyle{apsrev4-2}

\end{document}